\documentclass[letterpaper,twocolumn,prl,aps,superscriptaddress,amsmath,amssymb,floatfix]{revtex4-1}
\usepackage[latin9]{inputenc}
\usepackage{color}
\usepackage{verbatim}
\usepackage{float}
\usepackage{amsmath}
\usepackage{amssymb}
\usepackage{graphicx}
\usepackage{esint}

\usepackage[unicode=true,
 bookmarks=true,bookmarksnumbered=false,bookmarksopen=false,
 breaklinks=false,pdfborder={0 0 1},backref=false,colorlinks=true]
 {hyperref}
\hypersetup{
 linkcolor=magenta,urlcolor=blue,citecolor=blue,pdfstartview={FitH},hyperfootnotes=false}

\makeatletter

\usepackage{textcomp}
\usepackage{epstopdf}

\pdfpageheight\paperheight
\pdfpagewidth\paperwidth

\@ifundefined{textcolor}{}{%
 \definecolor{BLACK}{gray}{0}
 \definecolor{WHITE}{gray}{1}
 \definecolor{RED}{rgb}{1,0,0}
 \definecolor{GREEN}{rgb}{0,1,0}
 \definecolor{BLUE}{rgb}{0,0,1}
 \definecolor{CYAN}{cmyk}{1,0,0,0}
 \definecolor{MAGENTA}{cmyk}{0,1,0,0}
 \definecolor{YELLOW}{cmyk}{0,0,1,0}
}

\usepackage{xcolor}
\usepackage{soul}
\definecolor{blue}{rgb}{0,0,1}
\definecolor{red}{rgb}{1,0,0}
\definecolor{green}{rgb}{0,1,0}

\usepackage{soul}

\makeatother

\begin{document}

\address{School of Civil Engineering, Hefei University of Technology,
Hefei, Anhui 230009, P.R. China.}
\address{CAS Key Laboratory of Quantum Information, University of Science and Technology of China, Hefei, Anhui 230026, China.}
\address{Center for Quantum Information, Institute for Interdisciplinary
Information Sciences, Tsinghua University, Beijing 100084, China}
\address{CAS Center For Excellence in Quantum Information and Quantum Physics, University of Science and Technology of China, Hefei, Anhui 230026, China}
\address{National Laboratory of Solid State Microstructures, Nanjing University, Nanjing 210093, China.}

\title{Adiabatic conversion between gigahertz quasi-Rayleigh and quasi-Love modes for phononic integrated  circuits}

\author{Bao-Zhen Wang}
\address{School of Civil Engineering, Hefei University of Technology,
Hefei, Anhui 230009, P.R. China.}

\author{Xin-Biao Xu}
\email{xbxuphys@ustc.edu.cn}
\address{CAS Key Laboratory of Quantum Information, University of Science and Technology of China, Hefei, Anhui 230026, China.}
\address{CAS Center For Excellence in Quantum Information and Quantum Physics, University of Science and Technology of China, Hefei, Anhui 230026, China}

\author{Yan-Lei Zhang}
\address{CAS Key Laboratory of Quantum Information, University of Science and Technology of China, Hefei, Anhui 230026, China.}
\address{CAS Center For Excellence in Quantum Information and Quantum Physics, University of Science and Technology of China, Hefei, Anhui 230026, China}

\author{Weiting Wang}
\address{Center for Quantum Information, Institute for Interdisciplinary
Information Sciences, Tsinghua University, Beijing 100084, China}

\author{Luyan Sun}
\address{Center for Quantum Information, Institute for Interdisciplinary
Information Sciences, Tsinghua University, Beijing 100084, China}

\author{Guang-Can Guo}
\address{CAS Key Laboratory of Quantum Information, University of Science and Technology of China, Hefei, Anhui 230026, China.}
\address{CAS Center For Excellence in Quantum Information and Quantum Physics, University of Science and Technology of China, Hefei, Anhui 230026, China}
\author{Chang-Ling Zou}
\email{clzou321@ustc.edu.cn}
\address{CAS Key Laboratory of Quantum Information, University of Science and Technology of China, Hefei, Anhui 230026, China.}
\address{CAS Center For Excellence in Quantum Information and Quantum Physics, University of Science and Technology of China, Hefei, Anhui 230026, China}
\address{National Laboratory of Solid State Microstructures, Nanjing University, Nanjing 210093, China.}

\begin{abstract}
Unsuspended phononic integrated circuits have been proposed for on-chip acoustic information processing. Limited by the operation mechanism of a conventional interdigital transducer, the excitation of the quasi-Love mode in GaN-on-Sapphire is inefficient and thus a high-efficiency Rayleigh-to-Love mode converter is of great significance for future integrated phononic devices. Here, we propose a high-efficiency and robust phononic mode converter based on an adiabatic conversion mechanism. Utilizing the anisotropic elastic property of the substrate, the adiabatic mode converter is realized by a simple tapered phononic waveguide. A conversion efficiency exceeds $98\%$ with a $3\,\mathrm{dB}$ bandwidth of $1.7\,\mathrm{GHz}$ can be realized for phononic waveguides working at GHz frequency band, and excellent tolerance to the fabrication errors is also numerically validated. The device that we proposed can be useful in both classical and quantum phononic information processing, and the adiabatic mechanism could be generalized to other phononic device designs.
\end{abstract}
\maketitle

\section{Introduction}

In the past decades, phononic devices have played an important role in RF signal processing, and found applications in various fields include radars~\citep{Reindl2001}, communications~\citep{Campbell1998}, and sensing~\citep{Laenge2008,Liu2016}. Recently, phononic microstructures have also been used for the studies of quantum mechanics~\citep{Schuetz2015}, for the controlling of photons, electrons~\citep{Barnes2001,Hermelin2011}, quantum dots~\citep{Naber2006,Gell2008}, NV centers~\citep{Golter2016,Golter2016a}, and superconducting qubits~\citep{Gustafsson2014,Manenti2017,Chu2017,Satzinger2018}. In contrast to the conventional acoustic devices that utilize surface acoustic waves or bulk acoustic waves without lateral confinement of phonon, an unsuspended phononic integrated circuit (PnIC) platform has been proposed to guide and manipulate phonons with a strong lateral confinement~\citep{Fu2019,Wang2020}. In such a PnIC, phonons are confined by the high-acoustic-index-contrast between the waveguide and the substrate material, and various compact integrated phononic devices have been developed~\citep{Fu2019,Wang2020,Mayor2021,Shao2021}. Compared with the conventional acoustic devices, PnICs have the advantages of good extensibility, stability, and field restriction capability. Besides, by integrating phononic and superconducting devices together, a powerful hybrid integrated chip can be constructed for both classical and quantum information processing.

In a PnIC, the efficient excitation and manipulation of the phononic waveguide modes are essential. However, due to the anisotropy of the piezoelectric material, some phononic modes are piezoelectrically inactive. For instance, due to the limitation of the piezoelectric activity of the material and the fabrication capability, only the guiding mode with out-of-plane displacement (quasi-Rayleigh mode) could be efficiently excited by an interdigital transducer (IDT) fabricated on the GaN-on-Sapphire (GNOS) chip~\cite{Fu2019}. Hence, the quasi-Rayleigh mode has received more research attention and has been widely used in phononic devices. In fact, the quasi-Love mode that is dominated by in-plane displacements has different properties that are complementary to the quasi-Rayleigh mode, and thus has huge potential in future phononic applications. For example, the in-plane displacement of the quasi-Love mode makes the phonons have minimal damping in the air and liquid environment, which is favorable for high-quality integrated phononic devices. However, this mode is piezoelectrically inactive and is generally difficult to be directly excited by an IDT. Therefore, a robust and high-efficiency phononic mode converter between the quasi-Rayleigh and the quasi-Love modes is critical for a multi-functional PnIC.

Here, a high-efficiency mode conversion between the quasi-Rayleigh and the quasi-Love modes is investigated theoretically. The mode analysis of the GNOS waveguide structure is conducted to study the anisotropic-substrate-induced mode coupling of phononic modes. The acoustic wave propagation in the waveguide is analogous to the temporal state evolution in a quantum system, and a nearly perfect mode conversion can be realized by adiabatically changing the wave number along the waveguide, i.e. in a tapered phononic waveguide by adiabatically changing the waveguide width. As confirmed by 3D simulations, the proposed adiabatic converter has a large working bandwidth and is robust against geometry parameter uncertainties. We expect the devices that take advantage of the adiabatic evolution mechanisms would play a significant role in future studies of PnICs.

\begin{figure}
\begin{centering}
\includegraphics[width=1\columnwidth]{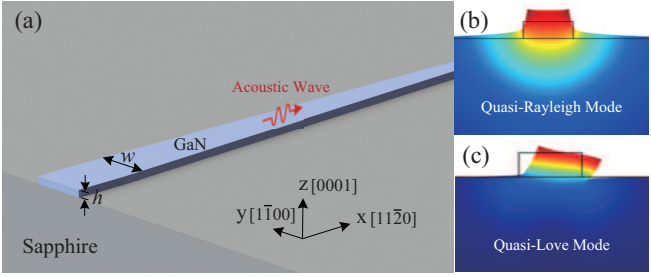}
\par\end{centering}
\caption{\label{fig:1}(a) Schematic of the strip GaN waveguide on a c-plane sapphire substrate along the $[11\bar{2}0]$ direction. $w$ and $h$ represent the width and thickness of the waveguide, respectively. (b)-(c) The displacement fields of the quasi-Rayleigh and  the quasi-Love eigenmodes at the cross-section of the phononic waveguide, respectively.}
\end{figure}

\section{anisotropic-substrate-induced mode coupling}
In this work, we consider a PnIC platform based on the GNOS~\cite{Fu2019}.
The schematic diagram of the model, which consists of a tapered GaN
waveguide on c-plane sapphire substrate, is shown in Fig.$\,$\ref{fig:1}(a).
The x, y, and z axes are consistent with the three orthogonal crystal
directions of the sapphire substrate: $[11\bar{2}0]$, $[1\bar{1}00]$,
and $[0001]$, respectively. In the following analyses, the thickness
of the GaN waveguide is set as $\mathrm{\mathit{h}=0.7\,\mu m}$.
Figures$\,$\ref{fig:1}(b) and (c) show the typical mode field distributions
of the fundamental quasi-Rayleigh ($R_{0}$) and quasi-Love ($L_{0}$)
modes of the phononic waveguide that are simulated by a finite-element method (COMSOL Multiphysics ver5.2). The quasi-Rayleigh and the quasi-Love modes are dominated by the out-plane and in-plane deformations, respectively. The exponential decay of the evanescent field in the substrate is similar to that of an optical waveguide~\cite{Saleh1991}, and the evanescent acoustic field enables the directional couplers for the coupling between phononic
waveguides and resonators~\citep{Wang2020}. The material parameters
of GaN and sapphire are given in Ref.~\citep{Pedros2005}.

\begin{figure}
\begin{centering}
\includegraphics[width=1\columnwidth]{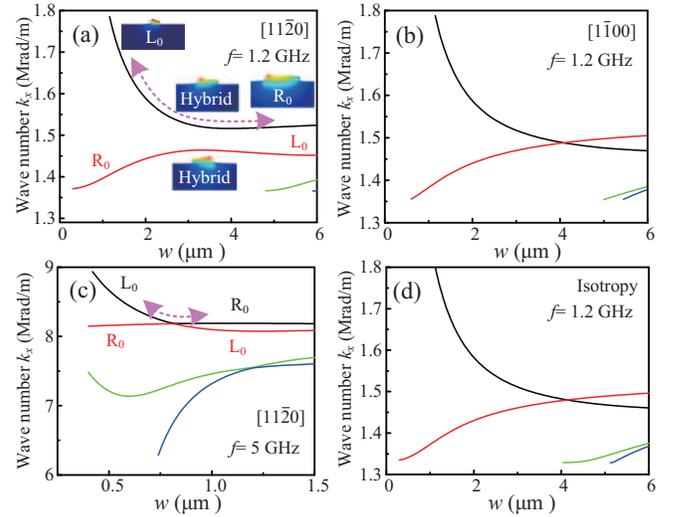}
\par\end{centering}
\caption{\label{fig:2}The calculated wave numbers as a function of $w$.
(a) the waveguide along the $[11\bar{2}0]$ direction with $f=1.2\,\mathrm{GHz}$. The insets show the deformation mode shapes at the specific widths.
(b) the waveguide along the $[1\bar{1}00]$ direction with $f=1.2\,\mathrm{GHz}$.
(c) the waveguide along the $[11\bar{2}0]$ direction with $f=5\,\mathrm{GHz}$.
(d) the waveguide on an isotropic substrate with $f=1.2\,\mathrm{GHz}$.}
\end{figure}

The propagation properties of the acoustic wave in the phononic waveguide
are closely related not only to the dimension and the elastic properties
of the waveguide, but also to the elastic properties of the substrate.
Because of its hexagonal crystal structure, GaN exhibits isotropic
mechanical properties in the c-plane and has five independent elastic
constants. However, the sapphire with a trigonal crystal structure
has six independent elastic constants and shows anisotropic elastic
properties. To see the influence of the anisotropy sapphire substrate
on the phononic modes in the GaN waveguide, we perform numerical
simulations and analyze the geometry-dependent mode wave vector, as
shown in Fig.$\,$\ref{fig:2}. At a given frequency of $1.2\,\mathrm{GHz}$,
the confined phononic modes in the waveguide are studied with the
waveguide width ranging from $0.5\,\mathrm{\mu m}$ to $6\,\mathrm{\mu m}$.

Figure$\,$\ref{fig:2}(a) investigates the dependence of the eigenmode
wave vector on the waveguide width for the waveguide along the
$[11\bar{2}0]$ direction. The inset plots the mode profiles of $R_{0}$
and $L_{0}$ modes. The black and red curves do not cross each other
and form an avoid-crossing line-shape in the figure. However, for the
modes in the avoid-crossing regions, the eigenmodes show hybridization
of horizontal and vertical vibrations and correspond to the coupling
between $R_{0}$ and $L_{0}$ modes. For a uniform waveguide, considering that
$R_{0}$ mode is excited in the avoid-crossing region, it can be decomposed as a superposition of two hybrid modes. Because of the difference
in the propagation speeds of the two hybrid modes, the output acoustic
field can be either $R_{0}$ or $L_{0}$ depending on the length of
the waveguide, which means the mode coupling between $R_{0}$ and $L_{0}$
is realized. Unlike the situation along the $[11\bar{2}0]$ direction,
the eigenmodes show distinct properties against $w$ for the waveguide
in the $[1\bar{1}00]$ direction, as shown in Fig.$\,$\ref{fig:2}(b). In this
direction, the black and red curves for $R_{0}$ and $L_{0}$ modes
could cross each other, which means that the coupling between the
two modes is negligible.

\begin{figure*}
\begin{centering}
\includegraphics[width=15cm]{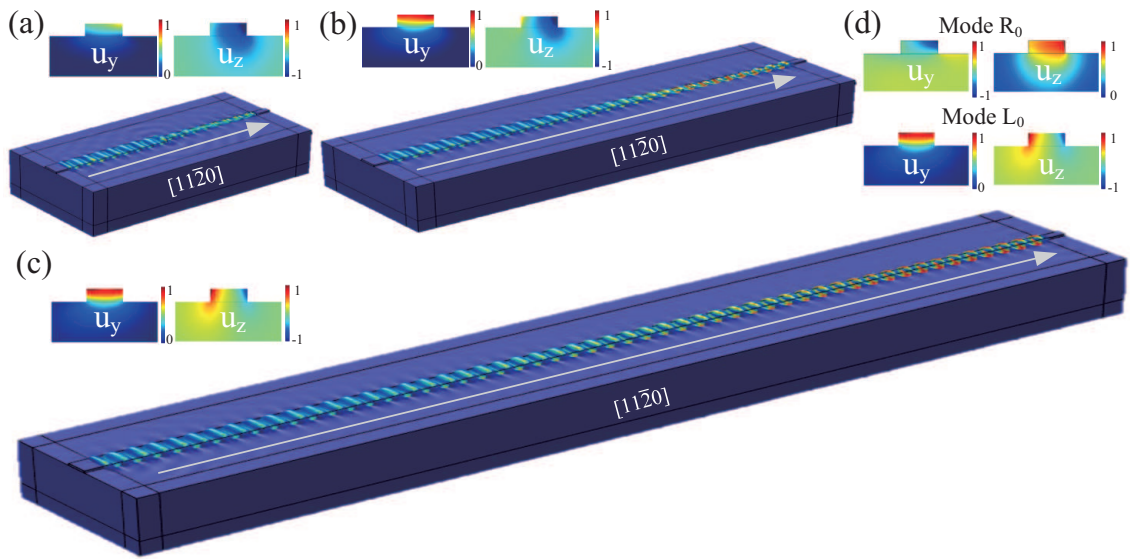}
\par\end{centering}
\caption{\label{fig:3} The simulated 3D displacement field distributions
of the acoustic wave propagating in the taper waveguide along the $[11\bar{2}0]$
direction and the $y$ and $z$ components of the output displacement
fields when the input field is mode $R_{0}$. (a) $l_{t}=0\,\mathrm{\mu m}$. (b)
$l_{t}=50\,\mathrm{\mu m}$. (c) $l_{t}=150\,\mathrm{\mu m}$. (d)
The eigenmodes of the tapered waveguide at the output port along the $[11\bar{2}0]$
direction.}
\end{figure*}

To further demonstrate that the mode coupling is mainly caused by
the anisotropic substrate, the phononic mode with a higher frequency
of $5\,\mathrm{GHz}$ along the $[11\bar{2}0]$ direction is also calculated
in Fig.$\,$\ref{fig:2}(c). Because the higher frequency the acoustic
wave has the shorter wavelength, the acoustic energy of the phononic
mode is mainly concentrated in the GaN layer so the influence of the
substrate anisotropy on the phonon propagation becomes weak. As depicted
in the curves, the gap between the black and red curves is reduced compared
to that in Fig.$\,$\ref{fig:2}(a), indicating the coupling strength
between $R_{0}$ and $L_{0}$ modes in the avoid-crossing region is
smaller. Furthermore, by changing the substrate as an isotropic material,
the phononic modes feature no coupling, as illustrated in Fig.$\,$\ref{fig:2}(d).
These results imply possible approaches to suppress the anisotropic
effect of the substrate by either changing the waveguide geometry
to avoid the mode hybridization or suppressing the evanescent field
in the substrate.

\section{Adiabatic mode conversion}

Compared to the mode coupling with a constant coupling strength, where
a proper waveguide length should be selected to realize the highest
mode conversion efficiency because of the periodic form of energy distribution
along the propagation direction, adiabatic mode conversion is another
widely used theory in quantum system and integrated photonic circuits. As a quantum control technique, the quantum adiabatic evolution has been extensively studied in coherent
control of quantum states and quantum computation~\citep{Farhi2000,Childs2001,Zhou2017}.
In an integrated optical waveguide, inspired by the similarity between
the evolution of the state in quantum mechanics and electromagnetic wave propagation
in optics, robust and broadband mode conversion with high-efficiency
has been realized by using the adiabatic theory on an integrated photonic
chip~\citep{Xu2019,Xu2019a,Kim2009,Dai2012}. The optical mode evolution
in waveguide is determined by the parameters of the waveguide. By
changing the parameters adiabatically in some situations, the adiabatic
mode conversion can be realized. For a tapered waveguide with a varying
waveguide width rate $dw/dx$, according to the Landau-Zener tunneling
theory~\citep{Xu2019}, the achievable mode conversion efficiency is
\begin{equation}
\eta=1-e^{-2\pi g^{2}/\left|\kappa\frac{dw}{dx}\right|},\label{eq:1}
\end{equation}
where $g$ is the mode coupling coefficient in the avoid-crossing
regions and $\kappa=d(\beta_{r}-\beta_{l})/dw$ with $\beta_{r}$ and
$\beta_{l}$ being the wave numbers of the quasi-Rayleigh and the quasi-Love
modes, respectively. Therefore, when the waveguide width changes adiabatically,
i.e. $dw/dx$ is small enough, a robust, broadband, and high-efficiency
mode converter can be realized.

\begin{figure*}[t]
\begin{centering}
\includegraphics[width=15cm]{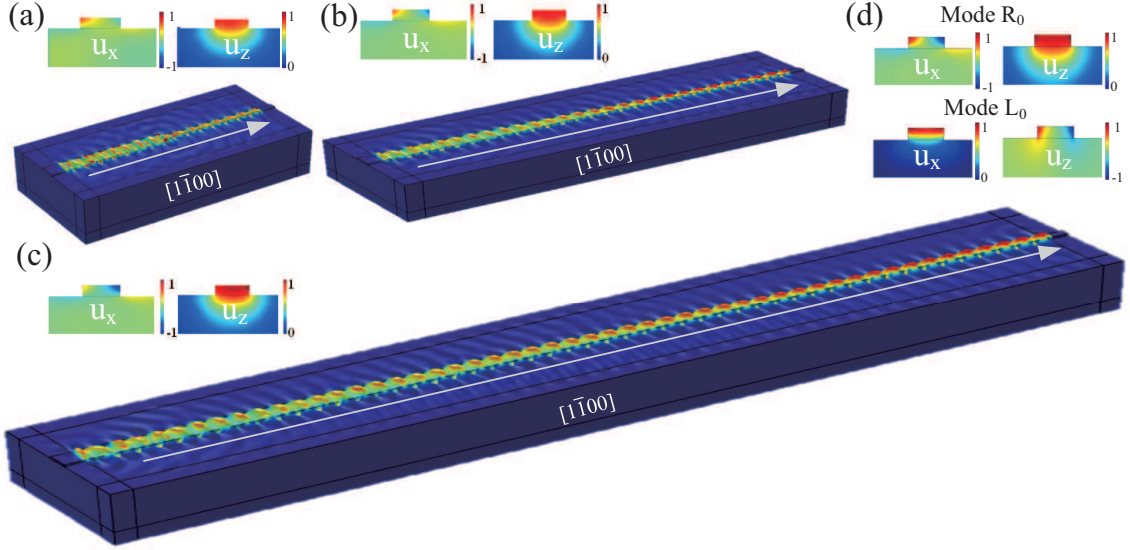}
\par\end{centering}
\caption{\label{fig:4} The simulated 3D displacement field distributions
of the acoustic wave propagating in the taper waveguides along the $[1\bar{1}00]$
direction and the $x$ and $z$ components of the output displacement
fields when the input field is mode $R_{0}$. (a) $l_{t}=0\,\mathrm{\mu m}$. (b)
$l_{t}=50\,\mathrm{\mu m}$. (c) $l_{t}=150\,\mathrm{\mu m}$. (d)
The eigenmodes of the tapered waveguide at the output port along the $[1\bar{1}00]$
direction.}
\end{figure*}

Following the pink dash curve with arrows in Fig.$\,$\ref{fig:2}(a),
the adiabatic mode conversion is numerically investigated by simulating
the propagation of the acoustic wave in a tapered phononic waveguide
along the $[11\bar{2}0]$ direction, as shown in Fig.$\,$\ref{fig:3}.
By fixing the input port waveguide width $w_{i}=5\,\mathrm{\mu m}$
and the output port waveguide width $w_{o}=2\,\mathrm{\mu m}$,
and varying the length of the tapered waveguide $l_{t}$, the evolution
of the displacement field and the corresponding $y$ and $z$
components of the output field profile for $l_{t}$= 0, 50 and $150\,\mathrm{\mu m}$
are illustrated in Figs.$\,$\ref{fig:3}(a)-(c), respectively. Here, the frequency
of the excited $R_{0}$ mode at the input port is $1.2\,\mathrm{GHz}$.
Fig.~\ref{fig:3}(d) shows the $y$ and $z$ components of the displacement
field of two eigenmodes ($R_{0}$ and $L_{0}$) of the output port
along the $[11\bar{2}0]$ direction. It can be found that the output fields,
when $l_{t}=150\,\mathrm{\mu m}$ as shown in Fig.$\,$\ref{fig:3}(c),
are very similar to those of the quasi-Love eigenmode of the output
port, which indicates a nearly-perfect conversion from the quasi-Rayleigh
mode to the quasi-Love mode. As expected, the converter becomes less adiabatic
since $dw/dx\propto1/l_{t}$ increases with the reduced taper
length so that the conversion efficiency decreases when $l_{t}=$ $0\,\mathrm{\mu m}$
and $50\,\mathrm{\mu m}$.

To verify the effect of the anisotropic substrate, we also simulate
the mode evolution in the tapered waveguide along the $[1\bar{1}00]$
direction, and the results are summarized in Fig.$\,$\ref{fig:4}.
Compared with the results in Fig.$\,$\ref{fig:3}, all the results
for the $[1\bar{1}00]$ direction show that the output mode is the quasi-Rayleigh
mode with $z$ component dominated displacement field, which indicates
the mode conversion almost does not occur. This behavior of the tapered
waveguides is consistent with our theoretical prediction, since the
substrate-induced coupling $g\approx0$ for the waveguide along the $[1\bar{1}00]$
direction, and $e^{-2\pi g^{2}/\left|\kappa\frac{dw}{dx}\right|}\approx1$
regardless of how adiabatic the waveguide width varies. Comparing
Figs.$\,$\ref{fig:4}(a), (b) and (c), there are still minor differences
in the displacement field profile at the output port. We attribute
these $l_{t}$-dependent effects to the varying waveguide geometry-induced coupling between the guided $R_{0}$ mode to other high-order
Rayleigh modes and diffused Rayleigh modes of the substrate.

Compared with $L_{0}$ mode, the $R_{0}$ mode could be more efficiently
excited by the IDT in GNOS, so it is of great interest to achieve the
efficient conversion from mode $R_{0}$ to mode $L_{0}$. To quantify
the performance of the converter, conversion efficiency $\eta$ is
calculated numerically. Since GaN is a piezoelectric
material with a relatively low piezo-coefficient and the velocity
of the acoustic wave is much smaller than the speed of light, the
quasi-static approximation is applicable for our model. Therefore,
only the elastic field is considered to calculate the power flow $P$
in the phononic waveguide as
\begin{equation}
P=\frac{1}{2}\mathrm{Re}\int(-\boldsymbol{V}^{*}\cdot\tilde{\boldsymbol{T}})\cdot d\vec{\boldsymbol{s}},\label{eq:2}
\end{equation}
where $V$ is the vector of velocity, $\tilde{\boldsymbol{T}}$ is
the stress field, the symbol ``*" means complex conjugate, and $\boldsymbol{s}$ is the cross-section of the phononic waveguide. Using the power flow, the mode conversion efficiency can be calculated according to the overlap integral of the displacement
field between the target eigenmode at the output port and the output
field. Here, the conversion efficiency from the quasi-Rayleigh mode to the
quasi-Love mode can be calculated as follows
\begin{align*}
\eta & =\frac{\left[\mathrm{Re}\int(\boldsymbol{V}_{Love}^{*}\cdot\tilde{\boldsymbol{T}}_{Taper}+\boldsymbol{V}_{Taper}^{*}\cdot\tilde{\boldsymbol{T}}_{Love}\right]/2\cdot d\vec{\boldsymbol{s}})^{2}}{\mathrm{Re}\int\boldsymbol{V}_{Taper}^{*}\cdot\tilde{\boldsymbol{T}}_{Taper}\cdot d\vec{\boldsymbol{s}}\cdot \mathrm{Re}\int\boldsymbol{V}_{Love}^{*}\cdot\tilde{\boldsymbol{T}}_{Love}\cdot d\vec{\boldsymbol{s}}},\label{eq:3}
\end{align*}
where the stress fields and the velocity fields are all obtained numerically, the subscript ``Taper" denotes the field distribution at the output port by simulating the evolution of the acoustic waves with a $R_{0}$ mode input, and the subscript ``Love" denotes the distribution for the $L_{0}$ mode of the output port.

\begin{figure}[t]
\begin{centering}
\includegraphics[width=1\columnwidth]{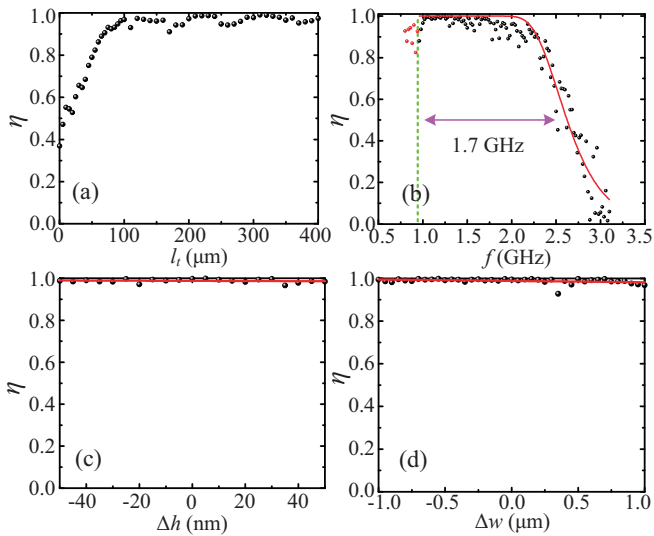}
\par\end{centering}
\caption{\label{fig:5}(a) The conversion efficiency $\eta$ from mode $R_{0}$ to
mode $L_{0}$ versus taper length at frequency $1.2\,\mathrm{GHz}$
(b) $\eta$ versus the frequency of phonons $f$ with the
taper length $l_{t}=220\,\mathrm{\mu m}$. (c)-(d) The tolerance of
the device to the variation of the thickness $h$ and the width $w$ of the waveguide,
respectively.}
\end{figure}

Figure$\,$\ref{fig:5} plots the results of conversion efficiency for
the tapered structure versus the geometry parameters. As shown in Fig.$\,$\ref{fig:5}(a),
for a fixed excitation frequency at $1.2\,\mathrm{GHz}$, the mode
conversion efficiency increases with the taper length $l_{t}$, and
saturates to unity when $l_{t}>100\,\mathrm{\mu m}$.
Such a saturation effect confirms the adiabatic conversion mechanism
as that the $\eta\approx1$ when the $dw/dz$ is small enough. The
robustness of the adiabatic conversion mechanism makes the structure
insensitive to the working frequency and also to the variations of
the geometry parameters. With a fixed taper length of $l_{t}=220\,\mathrm{\mu m}$, it is observed that $\eta\approx1$ in a wide range of frequencies
in Fig.$\,$\ref{fig:5}(b). The results indicate a broad working
bandwidth of about $1.7\,\mathrm{GHz}$ for $\eta>50\%$ in the frequency
range of $0.9\sim2.6\,\mathrm{GHz}$. Note that the lowest working
frequency at $0.9\,\mathrm{GHz}$ is limited by the cut-off frequency
of the waveguide at the output port, i.e. the $L_{0}$ mode does not
exist when frequency is below $0.9\,\mathrm{GHz}$. Besides, the robustness
of the adiabatic mode converter is further investigated by adding
a uniform change of thickness or width to the tapered waveguide at each position. The results
are summarized in Figs.$\,$\ref{fig:5}(c)-(d). As the waveguide thickness
varies within $\pm50\,\mathrm{nm}$ and the width of the waveguide
varies within $\pm1\,\mathrm{\mu m}$, $\eta$ is always larger
than 98\% and shows great tolerance against the geometry parameter
uncertainty in practical fabrications. The presented fluctuation of the conversion is mainly attributed to the numerical calculation accuracy.

\section{CONCLUSION}
In conclusion, we theoretically propose and investigate the phononic
adiabatic mode converter for PnICs. In a practical
PnIC platform, the conversion between the quasi-Rayleigh and the quasi-Love
modes is enabled by utilizing the anisotropy of the sapphire substrate.
Benefiting from the adiabatic mode evolution mechanism, a compact
tapered phononic waveguide allows efficient and robust conversion.
The mode conversion efficiency exceeding $98\%$ is obtained with a waveguide
length around $100\,\mathrm{\mu m}$. The device has a broad working
bandwidth of $1.7\,\mathrm{GHz}$ and is insensitive to the practical
fabrication uncertainties of the structure. Our work reveals the important
role of the anisotropy of the crystal compared to the isotropic materials,
and appeals careful design of integrated phononic devices by considering
the orientation of the device. Furthermore, our work represents one
example that shows the analogy between the acoustic wave propagation
dynamics and the evolution of the quantum system, and thus opens a new
avenue for designing useful phononic devices.

\begin{acknowledgments}
This work was supported by National Natural Science
Foundation of China (Grant No. 12061131011, 11922411, 11874342, 12104441, 92165209, and 11925404), the Natural Science Foundation of Anhui Provincial (Grant No. 2108085MA17 and 2108085MA22), China Postdoctoral Science Foundation (BX2021167), and Key-Area Research and Development Program of Guangdong Provice (Grant No. 2020B0303030001). This work was partially carried out at the USTC Center for Micro and Nanoscale Research and Fabrication. The numerical calculations in this paper have been done on the supercomputing system in the Supercomputing Center of University of Science and Technology of China.

\end{acknowledgments}



%

\end{document}